\documentclass[transmag]{IEEEtran}
\usepackage{latexsym}
\usepackage{graphicx}
\usepackage{amsfonts,amssymb,amsmath}
\usepackage{hyperref}

\usepackage{mathtools}
\usepackage{newtxmath}
\usepackage{algorithm}
\usepackage[caption=false,font=footnotesize]{subfig}
\usepackage{IEEEtrantools}
\usepackage{algorithmic}
\usepackage{dblfloatfix}
\usepackage{balance}
\usepackage{subfig}
\usepackage{acronym}
\usepackage{textcomp}

\input{acros}

\usepackage{prettyref}
\newrefformat{fig}{Fig.~\ref{#1}}
\newrefformat{conj}{Conjecture~\ref{#1}}
\newrefformat{tab}{Table~\ref{#1}}
\newrefformat{sec}{Section~\ref{#1}}
\newrefformat{subsec}{Section~\ref{#1}}
\newrefformat{subsubsec}{Section~\ref{#1}}

\usepackage{siunitx}
\DeclareSIUnit{\belmilliwatt}{Bm}
\DeclareSIUnit{\dBm}{\deci\belmilliwatt}
\DeclareSIQualifier{\isotropic}{i}
\DeclareSIQualifier{\carrier}{c}
\DeclareMathOperator*{\argmin}{\arg\,\min}

\newcommand{\diag}[1]{\operatorname{diag}\left({#1}\right)}
\newcommand*{\matr}[1]{\mathbf{#1}}

\makeatletter
\def\IEEElabelanchoreqn#1{\bgroup
\def\@currentlabel{\p@equation\theequation}\relax
\def\@currentHref{\@IEEEtheHrefequation}\label{#1}\relax
\Hy@raisedlink{\hyper@anchorstart{\@currentHref}}\relax
\Hy@raisedlink{\hyper@anchorend}\egroup}
\makeatother
\newcommand{\subnumberinglabel}[1]{\IEEEyesnumber
\IEEEyessubnumber*\IEEElabelanchoreqn{#1}}

\def\BibTeX{{\rm B\kern-.05em{\sc i\kern-.025em b}\kern-.08em T\kern-.1667em\lower.7ex\hbox{E}\kern-.125emX}}
\begin{document}

\title{Application-Based Coexistence of Different Waveforms on Non-Orthogonal Multiple Access}

\author{Mehmet Mert \c{S}ahin \IEEEmembership{Student Member, IEEE}, H{\"u}seyin Arslan
\IEEEmembership{Fellow, IEEE}
\thanks{The work of H. Arslan by the National Science Foundation
under Grant ECCS-1609581.}
\thanks{M. M. \c{S}ahin and H. Arslan are with the Department of Electrical Engineering, University of South Florida, Tampa, FL, 33620. H. Arslan is also with the Department of Electrical and Electronics Engineering, Istanbul Medipol University, Istanbul, TURKEY, 34810. E-mails: {\{mehmetmert, arslan\}}@usf.edu}
}

\IEEEtitleabstractindextext{

\begin{abstract}
The coexistence of different wireless communication systems such as LTE and Wi-Fi by sharing the unlicensed band is well studied in the literature. In these studies, various methods are proposed to support the coexistence of systems, including listen-before-talk mechanism, joint user association and resource allocation. However, in this study, the coexistence of different waveform structures in the same resource elements are studied under the theory of \ac{noma}. This study introduces a paradigm-shift on \ac{noma} towards the application-centric waveform coexistence. Throughout the paper, the coexistence of different waveforms is explained with two specific use cases, which are power-balanced \ac{noma} and joint radar-sensing and communication with \ac{noma}. In addition, some of the previous works in the literature regarding non-orthogonal waveform coexistence are reviewed. However, the concept is not limited to these use cases. With the rapid development of wireless technology, next-generation wireless systems are proposed to be flexible and hybrid, having different kinds of capabilities such as sensing, security, intelligence, control, and computing. Therefore, the concept of different waveforms' coexistence to meet these concerns are becoming impressive for researchers.    
\end{abstract}

\acresetall
}

\maketitle

\section{Introduction}

\IEEEPARstart{T}{he} idea of serving multiple users in the same wireless resources, including frequency, time, code, and space, has become an appealing research area over almost thirty years. Efforts to investigate new types of multiple access techniques under the constraint of scarce resources are named as multi-user detection and \ac{noma} for decades. The main motivation behind \ac{noma} having two different techniques, such as power-domain and code-domain, is the increased connectivity compared to \ac{oma}, which can meet the harsh requirements of the \ac{iot} \cite{NOMA_applications}. Several \ac{noma} schemes have been integrated into various standardization efforts. In LTE, the downlink \ac{noma} scheme, called multi-user superposition transmission (MUST), was studied for the 3rd Generation Partnership Project (3GPP) Release 14 \cite{3gpp.36.859}, whose motivation is weak in 5G, because higher performance gains can be provided with downlink massive \ac{mimo} \cite{Tian_2020_NOMA5G}. A study on the application of \ac{noma} for uplink transmission has been recently carried out for 3GPP Release 16, where different implementations of \ac{noma} have been studied \cite{3gpp.38.812}. However, since power-domain \ac{noma} has performance degradation in some cases, such as imperfect \ac{sic} and strict power control, it is not considered as a work-item in Release 17 \cite{Alouini_SurveyNOMA}. 

On the other hand, the transmit power of users is arranged in a way that the users' power received at the \ac{bs} is significantly different in order to enhance the overall system throughput in power-domain \ac{noma} \cite{PA_ULandDL_NOMA}. This arrangement in power-domain \ac{noma} introduces additional computational complexity that dynamically monitors the wireless system. Moreover, users transmitting at a similar power level may also be grouped in the case of high connectivity. Therefore, various researches have been conducted to find new \ac{noma} schemes that operate in power-balanced scenarios \cite{ncma_noma}. 

With rapid developments in hardware such as large antenna arrays for \ac{mmwave}, efficient amplifiers, and ultra-capable digital signal processing (DSP) chips; and software regarding algorithms for detection and estimation capabilities, radar and communication systems tend to intersect in order to provide efficient usage of radio resources. Combining these two different worlds to work in harmony will pave the way for new techniques in wireless technologies that may enable lots of promising applications that pave the way for new techniques in wireless technologies to emerge \cite{hanzo_2020}. For autonomous wireless networks, the capability to sense dynamically changing states of the environment and exchange information among various nodes needs to be integrated into 6G and beyond wireless systems \cite{chowdhury_6G_2020}. Also, radar-sensing is seen as an enabling technology for environment-aware communication in 6G \cite{bourdoux2020_6gWhitePaper}. Therefore, this trend encourages both industry and academia to plan and use the available radio resources efficiently. For example, the WLAN sensing group is organized under the IEEE 802.11 study group \cite{wlan_sensing_group}, where techniques to utilize the existing Wi-Fi frame as a sensing node are being developed regarding the future use cases of the concept. As a result, these concerns have created the joint radar-sensing and communication concept, which has gained a significant amount of attention from both industry and academia \cite{hanzo_2020,Rahman2020EnablingJC}.

The usage of \ac{mmwave} bands spurs the utilization of directed beams and larger bandwidths that improve sensing accuracy and data rate. So far, most of the papers investigate the optimal waveform to function jointly for both radar-sensing and communication, which is called joint radar-communication
(JRC) waveform \cite{HeathVorobyov2020}. The aim is to combine radar-sensing and communication into a single mmWave system that utilizes a standard waveform. This kind of system is aimed to be optimized regarding cost, size, power consumption, spectrum usage, and adoption of communication-capable vehicles, for example, in case of autonomous driving, which needs both radar and vehicle-to-vehicle (V2V) communication \cite{bliss_2017}. 

The coexistence of Wi-Fi systems with other wireless technologies is an important issue needed to be solved intelligently \cite{Doppler_2019_WiFiMag}. Various methods are proposed to cover the coexistence of systems, including listen-before-talk mechanism, joint user association and resource allocation \cite{Leung_CoexistenceLTEWiFi}. However, we introduce a novel concept on \ac{noma}, which is the coexistence of different waveform structures in the same resource elements. This paper studies two use cases of the introduced concept based on the superimposition of different waveforms non-orthogonally. The first use case is a power balanced \ac{noma} transmission scheme, which is presented by the authors in \cite{sahin_2020waveformdomain} where the OFDM and OFDM-IM waveforms are used in uplink transmission. Here, a downlink scenario of the proposed architecture is investigated with the calculation of achievable rates of the scheme. It is shown that when powers of users' signals are near to each other, the proposed scheme performs better compared to conventional power-domain \ac{noma} with only \ac{ofdm} waveform. The other use case is the implementation of the proposed concept in joint radar-sensing and communication.

Contributions of the paper can be listed as follows: 
\begin{itemize}
    \item A generalized method of different waveform coexistence with \ac{noma} is introduced for application-based wireless networks. 
    \item Previous work on the literature regarding different waveforms coexistence is overviewed. The works that are suitable for the \ac{noma} theory is included. 
    \item Two different use cases, which are power balanced \ac{noma} and joint radar-sensing and communication is studied in the view of waveform coexistence on the same wireless communication resources.  
    \item In the power balance \ac{noma} system, a novel transceiver design is proposed. \Ac{llr} calculations are done to perform the separability and detectability of the signals. 
\end{itemize}

The rest of the paper is organized as follows: \prettyref{sec:preWork} 
summarizes the literature work related to waveform coexistence on \ac{noma}. The power balanced \ac{noma} scheme, which is one of the use cases for waveform coexistence, is presented in \prettyref{sec:pb_NOMA}. \prettyref{sec:jrc_NOMA} introduces the use of waveform coexistence on \ac{noma} in the field of joint radar-sensing and communication. Future possible studies and conclusions are drawn in \prettyref{sec:conclusion}.

\section{Previous works} \label{sec:preWork}
\begin{table*}[!t]
	\renewcommand{\arraystretch}{1}
	\caption{Existing Literature Works on Waveform Coexistence }
	\label{tab:lit_review}
	\centering
	\begin{tabular}{|p{1.0in}|p{0.7in}|p{1.5in}|p{3.0in}|}
		\hline
		\bfseries Coexisted Waveforms &  \bfseries Reference & \bfseries Use case & \bfseries Comments\\
		\hline\hline
		OFDM + SC-FDMA & \cite{MBC_OFDMandSCFDMA} & It can be used in LTE heterogeneous networks with overloaded users. & Both waveforms are used in LTE system. A novel \ac{mud} approach is proposed to separate waveforms. \\
		\hline
		OFDM + CDMA & \cite{Sari_CDMAandOFDM} & It provides all users with the same data rate. & The proposed scheme relaxes the need for power allocation. \\
		\hline
		OFDM + OTFS & \cite{OTFS_NOMA} & It offers flexibility among users considering their mobility profile. & Compared to OFDM, OTFS has a different lattice structure, where symbols are placed in delay-Doppler plane. OTFS-NOMA allows high mobility user to spread its signal over the whole time-frequency plane. \\
		\hline
		OFDM + OFDM-IM & \cite{Tusha_2020,Dogan_2020_OFDM-IM_NOMA,sahin_2020waveformdomain} & It can be utilized in the power-balanced NOMA scenario and grant-free random access for URLLC & The OFDM-IM has some advantages over the OFDM such as ergodic achievable rate, PAPR reduction, and robustness to ICI \cite{IM_techniques}, Therefore NOMA with OFDM and OFDM-IM introduces more flexible multiple access scheme regarding user's requirements. \\
		\hline
		OFDM + FMCW & \cite{Sahin2020_MultifunctionalJRC} & It is designed for joint radar-sensing and communication functionality. & FMCW is a more suitable waveform for sensing purpose with condensed ambiguity function. The OFDM is utilized for communication purpose with high data-rate. \\
		\hline
		Multi-numerology OFDM-NOMA & \cite{Abusabah_2018_NOMAnumerology} & Overloaded 5G networks can benefit from multi-numerology OFDM-NOMA. & Mixed numerologies are utilized against the constraints of MUD, as well as reduced guard duration. \\ 
		\hline
		MUSA with different Hermite-Gaussian prototype filters &  \cite{catak_MUSA_2019} & It is proposed for 5G \ac{mmtc} service. & It increases the number of users by introducing third dimension, in addition to the time and frequency domains. In that dimension, symbols of users are spreaded by orthogonal Hermite-Gaussian functions. \\
		\hline
	\end{tabular}
\end{table*}

In this section, a few notable intelligent techniques proposed in the literature for the overlapping of users with different waveforms are presented. It is valuable to point out the definition of the waveform in wireless networks. The waveform consisting of symbol, pulse shape, and lattice is the physical shape of the signal carrying modulated information \cite{MulticarrierComm_Survey}. Any change in the physical shape of signal is considered as a different waveform throughout the paper. Therefore, the novel concept called coexistence of different waveforms proposes the superimposition of the signals with different physical shapes along over the non-orthogonal resources to introduce application-based flexibility, separability, and detectability. In \prettyref{tab:lit_review}, several previous works on different types of waveform coexistence are classified by pointing the aim behind such architectures.  

In \cite{Sari_CDMAandOFDM}, a scheme based on non-orthogonally coexisting \ac{ofdm} and \ac{cdma} is proposed. It is indicated that single-stage \ac{sic} achieves unsatisfactory performance under the presence of interference. Moreover, the power difference needed for \ac{sic} is compensated with the spreading nature of \ac{cdma}. The separability aspect on the overlapping of two different waveforms is investigated with an iterative receiver design that is computationally complex. 

Similarly, the coexistence of \ac{ofdm} and \ac{sc-fdma} is studied in \cite{MBC_OFDMandSCFDMA}. This work introduces a new degree of freedom to reuse the
occupied radio resources by intentionally creating co-existence between
different waveforms. It is shown that the proposed improved adaptive \ac{mud} approach utilizing iterative likelihood testing and \ac{sinr} based processing outperforms conventional \ac{sic}. 

Moreover, the \ac{otfs} waveform is used for the high mobility user, whereas the signal of the low mobility user is transmitted via the \ac{ofdm} waveform in \cite{OTFS_NOMA}. This \ac{otfs}-\ac{noma} concept provides flexibility among users regarding their mobility profile. Reference \cite{Ding_2020_OTFSNOMA_Beam} formulates the optimal beamforming design whose objective is to maximize the data rate of low-mobility NOMA users using \ac{ofdm}, with the constraint that the requirement for high-mobility users’ targeted data rate can be met.

Authors in \cite{catak_MUSA_2019} propose a novel \ac{musa} scheme with superimposing orthogonal waveforms whose filters are Hermite-Gaussian functions. These superimposed orthogonal functions introduce the third dimension in addition to time and frequency domains for each superimposed orthogonal functions. Here, the aim is to increase the total number of users that are served in the same resources. Besides high connectivity, the \ac{gfdm} waveform is proposed to induce lower latency.  

A \ac{noma} scheme is proposed for the multi-numerology \ac{ofdm} system in \cite{Abusabah_2018_NOMAnumerology}. In the multi-numerology \ac{ofdm}, which is common waveform family in 5G wireless networks, waveform structures differentiate with their lattice structure in time and frequency domain. The scheme utilizes the nature of mixed numerology \ac{ofdm} systems to reduce the constraints associated with the \ac{mud} operation.

A novel downlink \ac{noma} scheme with two users based on \ac{ofdm} and \ac{ofdmim} is proposed in \cite{Tusha_2020}, called OFDM-IM NOMA. In this scheme, the inherent power imbalance of \ac{ofdmim} leads to better throughput compared to conventional \ac{ofdm} \ac{noma} in the case of the power balanced scenario. It is shown that OFDM-IM NOMA outperforms the classical OFDM NOMA in terms of \ac{ber} under a total power constraint and achievable sum
rate. The system performance of IM-NOMA is based on the power difference between the overlapped users as well as the inherent features of the OFDM-IM signal. This scheme is studied as grant-free access for uplink transmission, where multiple users are overloaded in the same wireless communication resources \cite{Dogan_2020_OFDM-IM_NOMA}. It is proposed to reduce the effect of the collision on \ac{urllc} services.  Available resources are shared among the users when URLLC service is needed. K-repetitions transmission and \ac{mrc} receiver are utilized to
provide \ac{urllc} service. 

\section{Power-balanced NOMA transmission} \label{sec:pb_NOMA}
Here, the authors discussed the \ac{ber} performance of \ac{ofdm}-\ac{ofdmim} \ac{noma} in the downlink scenario. The proposed scheme includes \ac{ldpc}-aided soft interference cancellation that is discussed for the uplink scenario in \cite{sahin_2020waveformdomain}. A transceiver design utilizing \ac{ldpc} codes aided soft interference cancellation is presented to improve \ac{bler} performance when the received power of users are near to each other. The \ac{bler} performance can be improved through all received power level variations of users with more sophisticated coding schemes designed for the OFDM-IM waveform. 
\subsection{System Model}
\begin{figure}
\centerline{\includegraphics[width=3.5in]{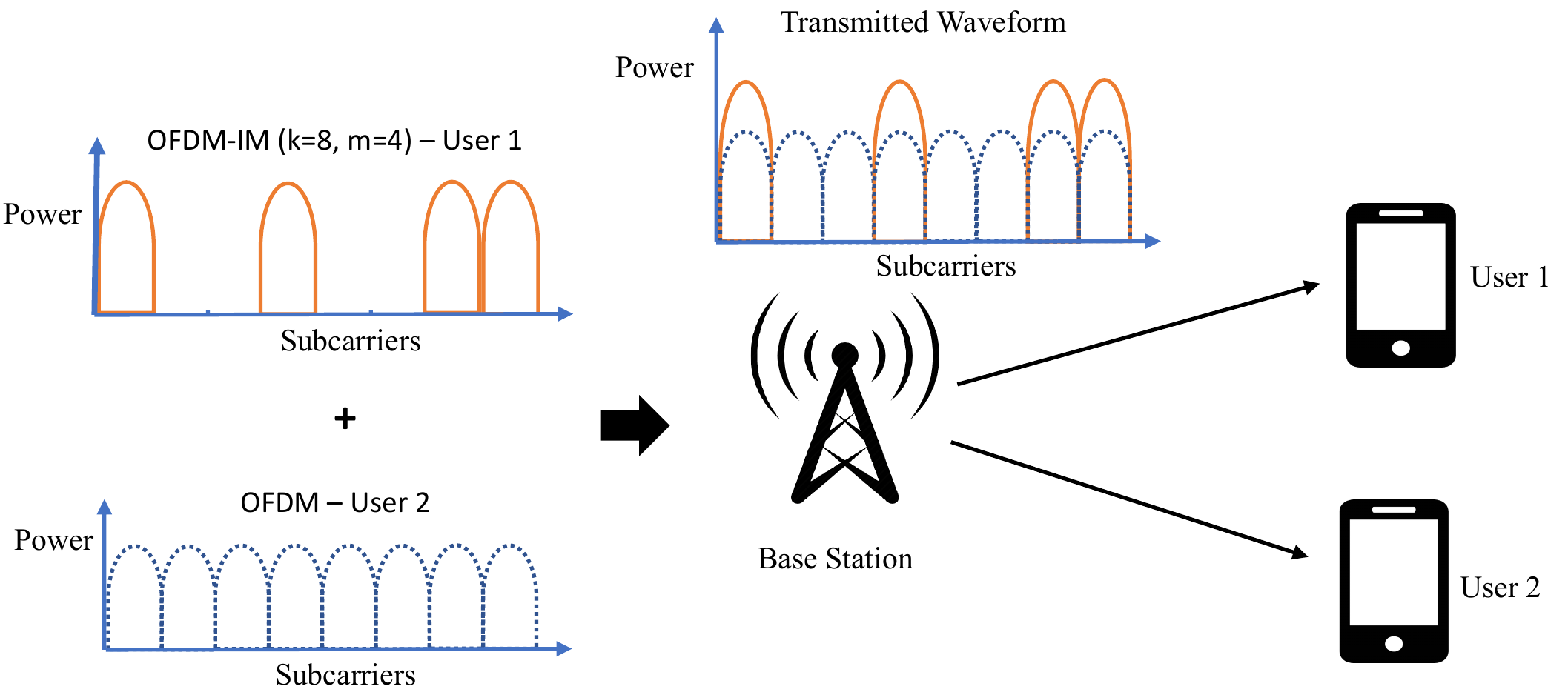}}
\caption{Two users uplink \ac{noma} scheme with \ac{ofdmim}+\ac{ofdm}.\label{fig:twoUserScheme}}
\end{figure}

For brevity, consider a two users downlink \ac{noma} scenario in \prettyref{fig:twoUserScheme}, where a single \ac{bs} transmits the superimposed signal to both users over $N$ subcarriers in the presence of frequency selective channels, including \ac{awgn}. The transceiver architecture of the proposed \ac{noma} scheme is presented in \prettyref{fig:wholeProcess}. The uplink scenario of the same scheme is investigated in the conference paper of the authors \cite{sahin_2020waveformdomain}.  

The \ac{llr} calculations are evaluated depending on the waveform type that is decoded first. It is shown that the proposed OFDM-IM and OFDM \ac{noma} scheme outperforms the conventional power-domain \ac{noma} scheme with OFDM waveform in terms of \ac{bler} performance in the power-balanced scenarios. Moreover, the proposed \ac{noma} scheme provides flexibility among users regarding their demands. 

\subsection{Conventional power-domain NOMA with OFDM+OFDM \label{sec:ofdm_ofdm}} 
Firstly, \ac{bs} encodes messages of \acp{ue} by \ac{ldpc} codes and modulate via \ac{qam}, where data symbols of users are drawn from a complex symbol alphabet $\mathbb{S}$. Then, these symbols are \ac{ofdm} modulated and transmitted to be received by each \ac{ue} over the same \ac{re}. Moreover, $p_1$ and $p_2$ denote the signal power of user 1 and user 2 for each subcarrier, respectively. In the \ac{ofdm} with total $N$ subcarriers, the total powers of user 1 and user 2 become $P_{1} = Np_{1}$ and $P_{2} = Np_{2}$, respectively. After the process of \ac{fft} and removal of cyclic prefix, the baseband received signal at the $n$th subcarrier for each \ac{ue} is expressed as follows:	
\begin{IEEEeqnarray}{rCl} 
\subnumberinglabel{eq:ofdm_ofdmNOMA}
r_{1,n} & = & h_{1,n} \left(\sqrt{p_1} u_{1,n} + \sqrt{p_2} u_{2,n}\right) + w_n, \IEEEeqnarraynumspace  \label{eq:receivedSampled} \\
\noalign{\noindent and
\vspace{2\jot}}
r_{2,n} & = & h_{2,n} \left(\sqrt{p_1} u_{1,n} + \sqrt{p_2} u_{2,n}\right) + w_n, \IEEEeqnarraynumspace  \label{eq:receivedCoeff}
\end{IEEEeqnarray}
where $h_{1,n}$, $h_{2,n}$, $u_{1,n}$, and $u_{2,n}$ are the channel gains and data symbols of users 1, and 2, respectively. Also, $w_n\sim\mathcal{CN}\left(0,\sigma^2\right)$ denotes the \ac{awgn} at the $n$th subcarrier.
	
Assuming that the signal of user 1 is decoded first where user 1 has more power, the capacity of user 1 ($R_1$) in conventional power-domain \ac{noma} is given by	
\begin{IEEEeqnarray}{rCl}
	R_1 &=& \sum_{n=1}^N\log_2\left(1+\frac{p_{1}h_{1,n}}{\sigma^2+p_{2}h_{1,n}}\right) 
	\text{bit/s/Hz.}
	\label{eq:dataRateOFDMFirst}
\end{IEEEeqnarray} 
Assuming perfect \ac{sic}, which is infeasible, the capacity of user 2 ($R_2$) is calculated as follows: 	
\begin{IEEEeqnarray}{rCl} 
	R_2 &=& \sum_{n=1}^N\log_2\left(1+\frac{p_{2}h_{2,n}}{\sigma^2}\right) 
	\text{bit/s/Hz.}
	\label{eq:dataRateOFDMSecond}
\end{IEEEeqnarray}
In the case of \ac{mlmud} without \ac{sic}, the decoding order does not have any effect on the sum-rate; therefore, any arbitrary decoding order can be assumed to be performed \cite{uplink_NOMA}. On the other hand, when \ac{ml}-\ac{mud} with \ac{sic} is used, the user with higher received power should be decoded first. 	
\subsection{NOMA with OFDM-IM+OFDM \label{sec:ofdm_ofdm-im}}	
\begin{figure}
	\subfloat[]{
		\includegraphics[width=3.5in]{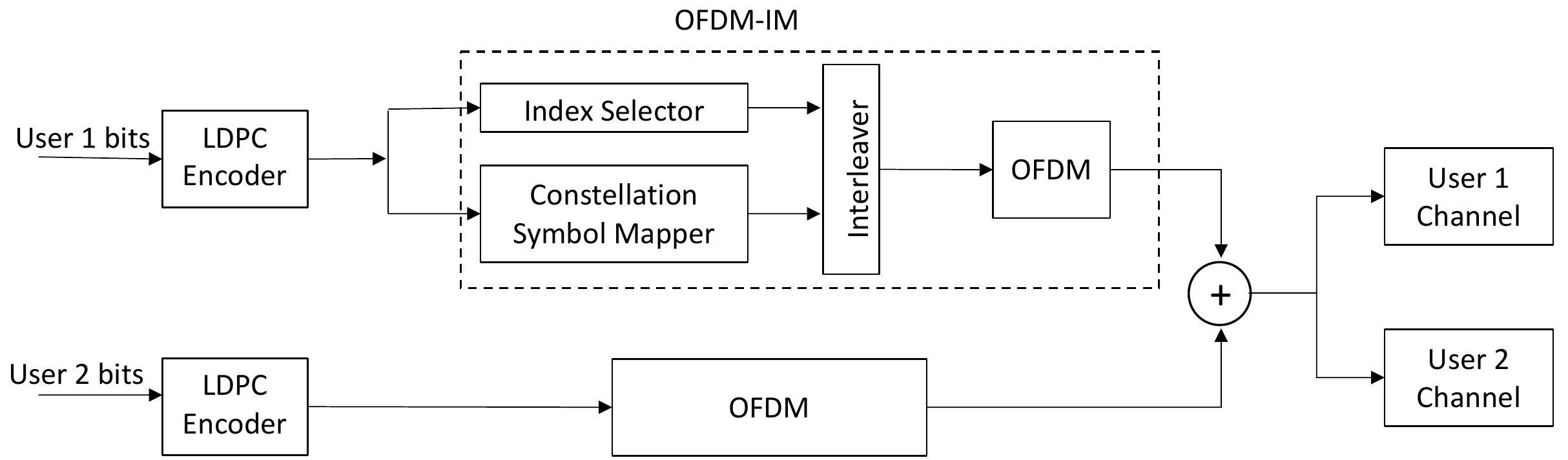}
		\label{fig:gen_frame1}}\hfil
	\subfloat[]{
		\includegraphics[width=3.5in]{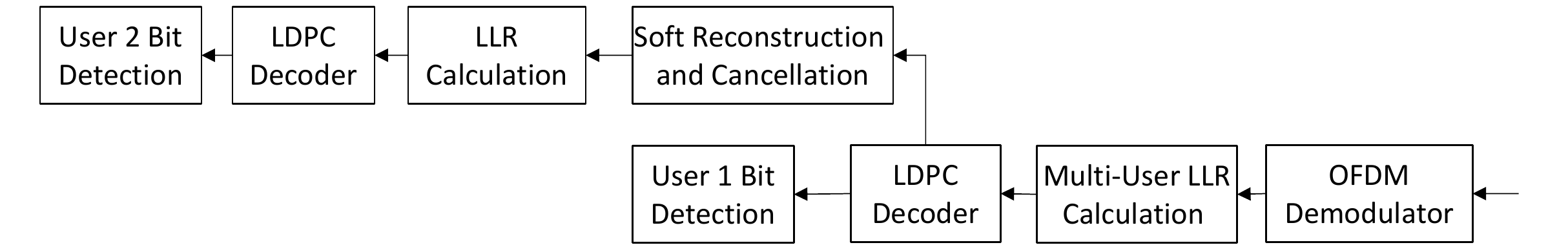}
		\label{fig:gen_frame2}}\hfil
	\caption{Proposed transmission and reception scheme (a) Coding and modulating of user 1 (\ac{ofdmim}) and user 2 (\ac{ofdm}) signals, (b) Demodulating and decoding of the superimposed received signal with \ac{ldpc} codes aided soft reconstruction and cancellation.}
	\label{fig:wholeProcess}
\end{figure}

As \prettyref{fig:twoUserScheme} depicts, \ac{ofdmim} waveform has been utilized for user 1, whereas \ac{ofdm} waveform is used to send the signal of user 2 over $N$ subcarriers. In the \ac{ofdmim} scheme \cite{ofdm_IM}, the total $Q=Q_1+Q_2$ bits are transmitted as follows: Firstly, $N$ subcarriers are split into total $G$ subblocks consisting of $k$ subcarriers. The $Q_1$ bits are used to determine the indices of $m$ active subcarriers where the total number of active subcarrier positions is denoted as $c = Gm$. In each subblock $\beta$, only $m$ out of $k$ subcarriers have activated.  
Activated subcarriers are used to map $Q_2$ bits on to $M$-ary signal constellation symbols selected from the complex set $\mathbb{S}$. The information of user 1, which is carried in the subblock $\beta$, is given by $\matr{u_{1,\beta}} = \left[u_{1,\beta}^{(1)} \ldots u_{1,\beta}^{(Q)}\right]$. Let $\Omega^{\beta}$ denote the set of active subcarrier indices, whereas $\Bar{\Omega}^{\beta}$ is the complement of it, including empty subcarriers at the $\beta$-th subcarrier. As seen in \prettyref{fig:gen_frame1}, the interleaved grouping is performed to increase the achievable rate of \ac{ofdmim} \cite{rate_OFDM-IM}. In subblock $\beta$, the vector of modulated symbols of user 2 carried with \ac{ofdm} waveform is denoted by $\matr{u_{2,\beta}} = \left[u_{2,\beta}^{(1)} \ldots u_{2,\beta}^{(k)}\right]$.	
	
After \ac{fft} and cyclic prefix removal, the superimposed received signal for the users at the $n$th subcarrier becomes   
	\begin{IEEEeqnarray}{rCl} \label{eq:sysModel}
	r_{1,n} &=& h_{1,n} \left(\sqrt{\frac{k p_1}{m N}} u_{1,n} + \sqrt{p_2} u_{2,n}\right) + w_{1,n}, 
	\\
	r_{2,n} &=& h_{2,n} \left(\sqrt{\frac{k p_1}{m N}}  u_{1,n} + \sqrt{p_2} u_{2,n}\right) +   w_{2,n}, 
	\end{IEEEeqnarray}
where $ u_{1,n} \in \mathbb{S^\prime}=\{0,\mathbb{S}\}$. Moreover, denote $\matr{r}_{\beta} \in \mathbb{C}^{1\times k}$ as the received signal at the $\beta$th subgroup. \prettyref{fig:gen_frame2} depicts the reception process of the proposed \ac{noma} scheme by decoding the \ac{ofdmim} waveform first. However, the decoding may not always start with the \ac{ofdmim} waveform. It depends on both power, subcarrier allocation, and modulation order.
Here, calculations for the capacity of users are performed on one subblock group. If the \ac{ofdm} waveform is removed in the superimposed signal and assuming that the subcarriers in the subgroup are faded independently, the capacity of user 2 is written as
\begin{IEEEeqnarray}{rCl} 
	R_2 &=& \sum_{n \in \Bar{\Omega}^{\beta}}\log_2\left(1+\frac{p_{2}h_{2,n}}{\sigma^2} \right) + \sum_{n \in \Omega^\beta}\log_2\left(1+\frac{p_{2}h_{2,n}}{\sigma^2 + \frac{k P_1}{m N}h_{2,n}} \right), \IEEEeqnarraynumspace  
	\label{eq:dataRateOFDMSecond}
\end{IEEEeqnarray}	
After successfully removing the \ac{ofdm} signal from the superimposed signal, the capacity of user 1 with interleaved OFDM-IM waveform is lower bounded as follows \cite{rate_OFDM-IM}:
\begin{IEEEeqnarray} {rCl}
    R_1 &=& \frac{m}{k} \log_2(M) + \frac{\log_2 \left(C(k,m)\right)}{k} \nonumber \\ && - \frac{1}{C(k,m) k M^{m}} \nonumber \\ 
    && \times \sum_{j=1}^{C(k,m)} \sum_{p^{m}} \times \log_2\left(\sum_{j^\prime = 1}^{C(k,m)}\sum_{p^{\prime m}} \frac{1}{\det\left(\mathbf{I}_k + \mathbf{\Lambda}_{j,j^\prime}\right)}\right)
    \label{eq:ofdmIM_rate}
\end{IEEEeqnarray}   
where $C(k,m)$ denotes the binomial coefficient, $\mathbf{I}_k$ is the $k \times k$ unit matrix, $\mathbf{\Lambda}_{j,j^\prime}$ is an $k \times k$ diagonal matrix whose $i$-th diagonal element is given as 
\begin{equation}
  [\mathbf{\Lambda}_{j,j^\prime}]_{i,i} = \left\{ \,
\begin{IEEEeqnarraybox}[][c]{l?s}
\IEEEstrut
\frac{p_1 k}{2 \sigma^2 m} \left|{s}_{p_{\Omega_j^{-1}(i)}} - {s}_{p^\prime_{\Omega_j^{-1}(i)}} \right|^2, &  $i \in \Omega_j \cap \Omega_{j^\prime}$, \\
\frac{p_1 k}{2 \sigma^2 m} \left|{s}_{p_{\Omega_j^{-1}(i)}}\right|^2, &   $i \in \Omega_j \cap \Bar{\Omega}_{j^\prime}$, \\
\frac{p_1 k}{2 \sigma^2 m} \left| {s}_{p^\prime_{\Omega_j^{-1}(i)}} \right|^2, &   $i \in \Bar{\Omega}_j \cap \Omega_{j^\prime}$, \\
0, &   $i \in \Bar{\Omega}_j \bigcap \Bar{\Omega}_{j^\prime}$,
\IEEEstrut
\end{IEEEeqnarraybox}
\right.  
\end{equation}
where $\mathbf{s}$ denotes the QAM modulated symbols in the activated $m$ subcarriers, $\mathbf{s} = \left[s_{p_1},\ldots,s_{p_m}\right] \in \mathbb{S}^{m}$. As mentioned in \cite{rate_OFDM-IM}, it should be noted that if $\Omega_j(r) = i$, then $\Omega_j^{-1}(i) = r \cdot \sum_{p^(n)} = \sum_{p_1 = 1}^M \ldots \sum_{p_n=1}^M$. To conclude, \prettyref{eq:ofdmIM_rate} is the lower bound of the achievable data rate with interleaved OFDM-IM waveform for user 1.

\subsection{LLR Calculations}
This section includes the \ac{llr} calculations of each user for two different \ac{noma} schemes. Calculated \acp{llr} are sent to the \ac{ldpc} decoder as input. For the sake of fair comparison, we have used the log-sum approximation technique \cite{BICM} to calculate approximate \acp{llr} of two different \ac{noma} schemes.
\subsubsection{LLR calculations for NOMA with OFDM+OFDM}	
With \ac{ml}-\ac{mud} algorithm, the \ac{llr} of the bit $i$ of user 1 at the $n$th subcarrier, $\Lambda_{n^{(i)}}^{u_1}$, is calculated as
\begin{IEEEeqnarray}{rCl} \label{LLR_OFDM_OFDM}
	\Lambda_{n^{(i)}}^{u_1}&=&\log \left( \frac{f(r_n | u_{1,n}^{(i)}=0)}{f(r_n | u_{1,n}^{(i)}=1)} \right) \nonumber \\
	& \approx & \min_{u_{1,n} \colon u_{1,n}^{(i)}\in \mathbb{S}_{1}^i,\ u_{2,n}\in \mathbb{S}} \frac{\|r_n-h_{1,n} (u_{1,n}-u_{2,n}) \|^2}{\sigma^2}  \nonumber \\ 
	&& - \min_{u_{1,n} \colon u_{1,n}^{(i)}\in \mathbb{S}_{0}^i, \ u_{2,n}\in \mathbb{S}} \frac{\| r_n-h_{1,n} (u_{1,n}-u_{2,n}) \|^2}{\sigma^2},\IEEEeqnarraynumspace 
	\end{IEEEeqnarray} 
	where $\mathbb{S}_{b}^i \subset \mathbb{S}$ denotes the set of all symbols $\alpha \in \mathbb{S}$ whose label has $b \in \{0,1\}$ in bit position $i$. The complexity of this \ac{llr} calculation, in terms of complex multiplications, becomes $\sim \mathcal{O}\left(\left|\mathbb{S}\right|^{2}\right)$.
	After \ac{ldpc} decoder is fed with \acp{llr}, the symbols of user 1 is reconstructed and subtracted from the superimposed signal with inevitable \ac{sic} error. The \acp{llr} of user 2 are calculated with the remaining signal and sent to the \ac{ldpc} decoder in order to obtain bit decisions of user 2. 
	
	\subsubsection{\ac{llr} calculations for \ac{noma} with \ac{ofdmim}+\ac{ofdm}}
	
	The \ac{llr} calculations for users' bits in the \ac{ofdmim}+\ac{ofdm} \ac{noma} scheme depend on which waveform is decided to be decoded first. As it is shown via numerical results in \prettyref{sec:numEval_1}, the total power level is not the unique limitation to decide which waveform should be decoded first. By decoding the \ac{ofdmim} waveform first, the \ac{llr} of the bit $i$ of user 1 at the $\beta$th subgroup, $\Lambda_{\beta^{(i)}}^{u_1}$ is 
	\begin{IEEEeqnarray}{rCl} 
		\Lambda_{\beta^{(i)}}^{u_1}
		&=&\log \left( \frac{f(\matr{r}_{\beta} | u_{1,\beta}^{(i)}=0)}{f(\matr{r}_{\beta} | u_{1,\beta}^{(i)}=1)} \right)  \nonumber\\
		&\approx& \min_{\matr{u}_{1,\beta} \colon u_{1,\beta}^{(i)}=1,\ \matr{u}_{2,\beta} \in \{S\}^k} \frac{\| \matr{r}_\beta-\matr{h}_{1,\beta}\odot  (\matr{u_{1,\beta}}-\matr{u}_{2,\beta}) \|^2}{\sigma^2}  \nonumber \\ &&
		\IEEEeqnarraymulticol{1}{r}{
		 -\>\min_{\matr{u}_{1,\beta} \colon u_{1,\beta}^{(i)}=0,\matr{u}_{2,\beta} \in \{S\}^k} \frac{\| \matr{r}_\beta-\matr{h}_{1,\beta}\odot (\matr{u_{1,\beta}}- \matr{u}_{2,\beta})\|^2}{\sigma^2}\IEEEeqnarraynumspace} 
	\end{IEEEeqnarray} 	
	where $\matr{h}_{1,\beta} \in \mathbb{C}^{1\times k}$ and $\matr{h}_{2,\beta} \in \mathbb{C}^{1\times k}$ denote the \ac{csi} of users 1 and 2 through $\beta$th subgroup, respectively, and $\odot$ denotes Hadamard multiplication.
	When the \ac{ofdmim} waveform is decoded first, the complexity of \ac{llr} calculation, in terms of complex multiplications, becomes $\sim\mathcal{O}\left(c\left|\mathbb{S}\right|^{m} \left|\mathbb{S}\right|^{k}\right)$. On the other hand, starting the decoding process with the \ac{ofdm} waveform, the \ac{llr} of the bit $i$ of user 2 at the $n$th subcarrier, $\Lambda_{n^{(i)}}^{u_2}$, becomes	
	\begin{IEEEeqnarray}{rCl} \label{LLR_OFDM_IM}
		\Lambda_{n^{(i)}}^{u_2} & = &\log \left( \frac{f(r_n | u_{2,n}^{(i)}=0)}{f(r_n | u_{2,n}^{(i)}=1)} \right) \nonumber \\
		& \approx & \min_{u_{2,n} \colon u_{2,n}^{(i)}\in \mathbb{S}_{1}^i,\ u_{1,n}\in \mathbb{S}^{\prime}} \frac{\|r_n-h_{2,n} (u_{2,n}-u_{1,n}) \|^2}{\sigma^2} \nonumber \\ && -  \min_{u_{2,n} \colon u_{2,n}^{(i)}\in \mathbb{S}_{0}^i, \ u_{1,n}\in \mathbb{S}^{\prime}} \frac{\| r_n-h_{2,n}(u_{2,n}-u_{1,n}) \|^2}{\sigma^2}. \IEEEeqnarraynumspace 
	\end{IEEEeqnarray} 
	By decoding the \ac{ofdm} waveform first, the complexity of the \ac{llr} calculation, in terms of complex multiplications, becomes $\sim\mathcal{O}\left(\left|\mathbb{S^{\prime}}\right|\left|\mathbb{S}\right|\right)$. The waveform, whichever is decoded first, is reconstructed and subtracted from the aggregate received signal before the next user's signal is decoded.
	
\begin{figure}
\subfloat[]{
    \centerline{\includegraphics[width=3.5in]{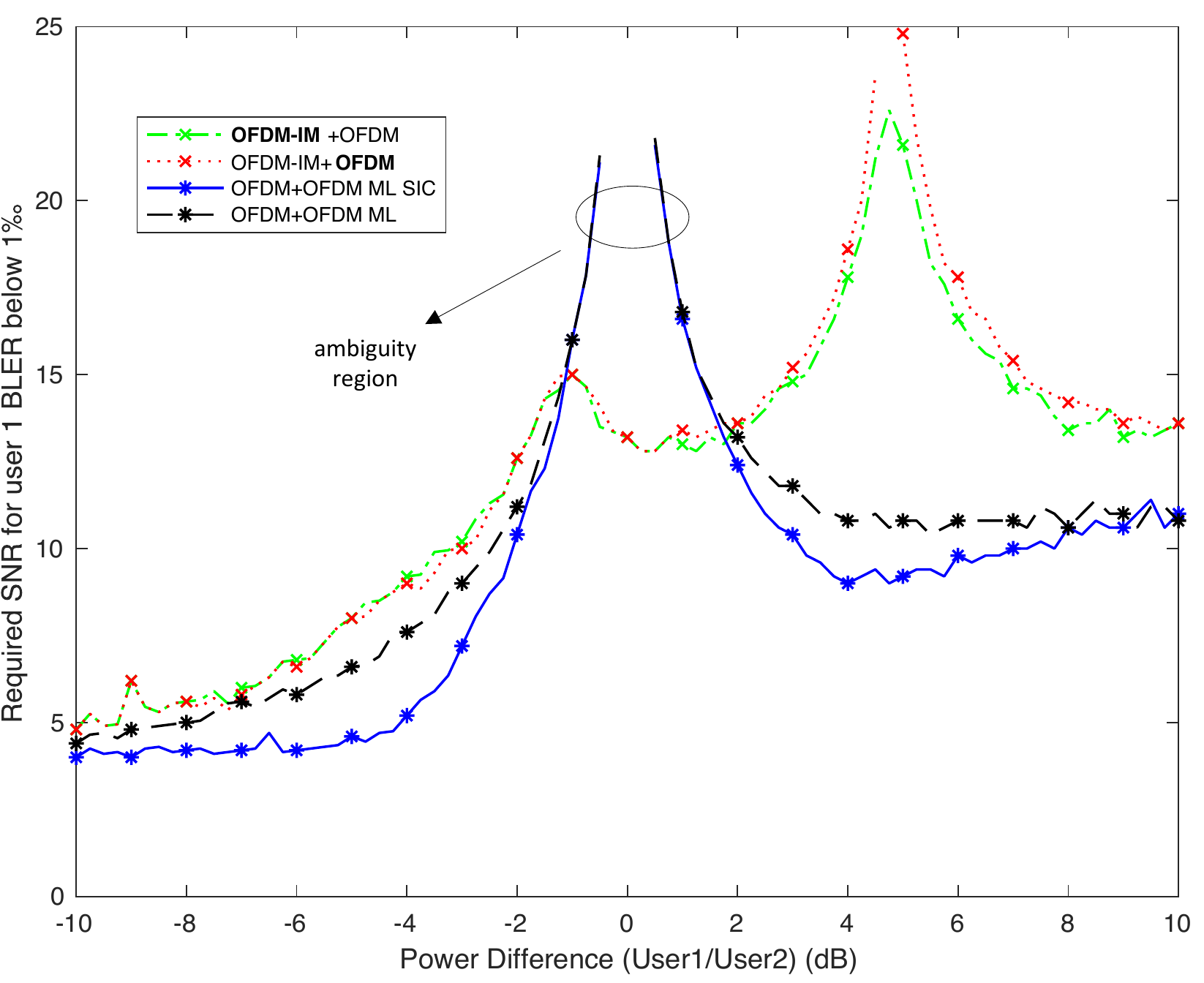}}
    \label{fig:bler1} }
\newline
\subfloat[]{
  \centerline{\includegraphics[width=3.5in]{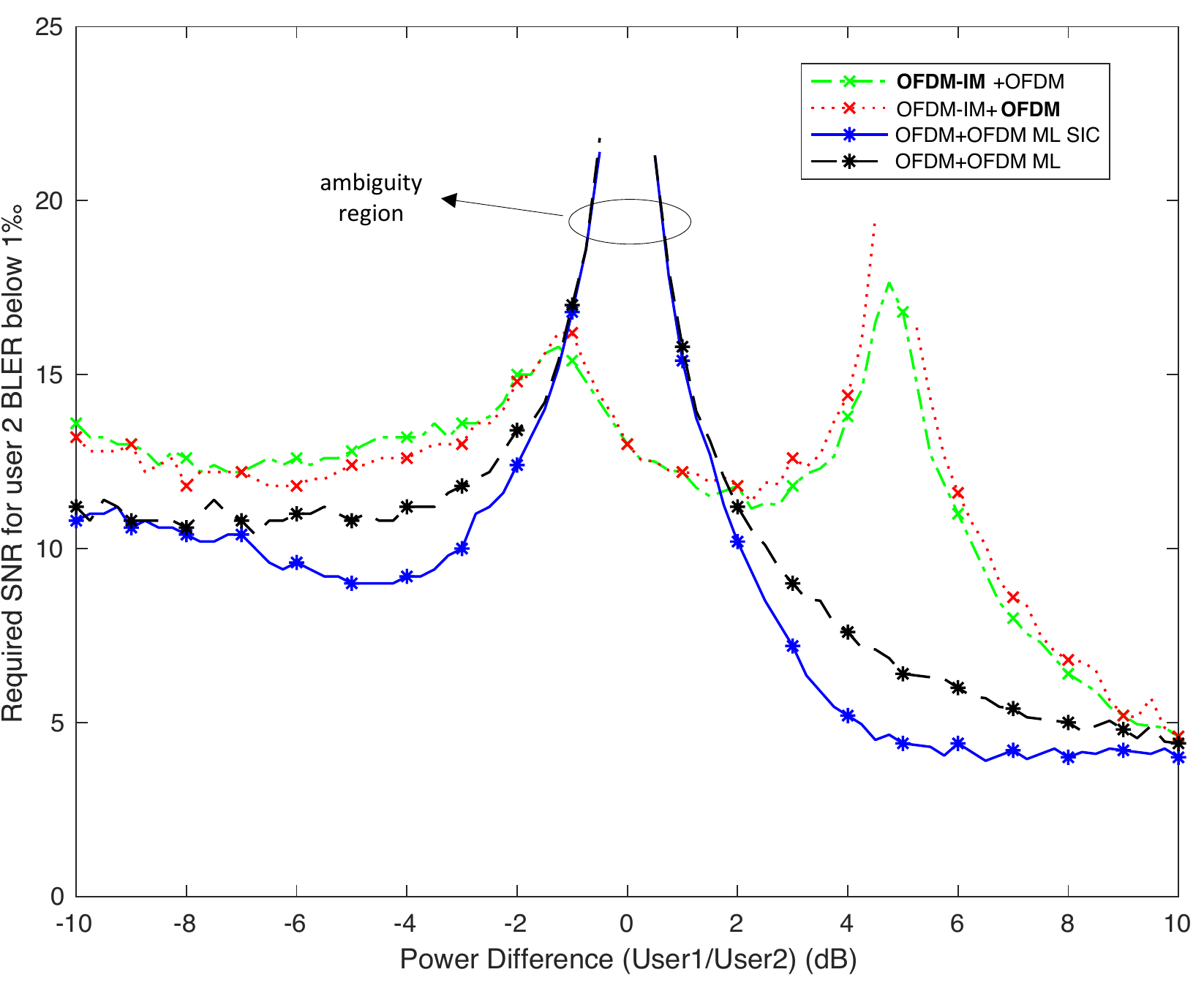}}
\label{fig:bler2}
}
\caption{Comparison of proposed waveform-domain \ac{noma} and conventional \ac{ofdm}-\ac{ofdm} \ac{noma} in downlink transmission, (a) BLER of user 1, (b) BLER of user 2.}
\label{fig:blerPlot}
\end{figure}

\subsection{Numerical Evaluation} \label{sec:numEval_1}
The proposed technique is evaluated numerically through Monte Carlo simulations. As in \cite{Sari_CDMAandOFDM}, the performance of the proposed \ac{noma} scheme is evaluated with \ac{bler} metric. Since the \ac{sic} is not perfectly performed in practice, achievable rate analysis under the perfect \ac{sic} condition misleads the comparison of the schemes. As for the modulation order, QPSK signaling is used for both \ac{noma} schemes, where equal data rate is satisfied with three active subcarriers in groups of four subcarriers ($m=3$, $k=4$) for the user utilizing the \ac{ofdmim} waveform. For \ac{ofdm}+\ac{ofdm} \ac{noma}, the user with high received power is decoded first, then reconstructed, and eliminated from the superimposed signal. On the other hand, for \ac{ofdmim}+\ac{ofdm} \ac{noma}, the decoding order is determined according to waveform type. Firstly decoded waveform is shown as bold for all given plots. 
	
\prettyref{fig:bler1} and \prettyref{fig:bler2} demonstrate the performance of conventional \ac{ofdm} \ac{noma} and proposed \ac{ofdmim} \ac{noma} schemes over the frequency selective channel with 10 taps for user 1 and user 2, respectively. Code rate is selected as 0.5 with 256 block length. Throughout the simulation, it is assumed that channel knowledge is present at the receiver.
		
The vertical axis denotes the required \ac{snr} for a user to achieve the target \ac{bler} of $1\text{\textperthousand}$, whereas the horizontal axis denotes the power difference in terms of \si{dB} between two different users. \ac{ofdmim}+\ac{ofdm} \ac{noma} is superior in terms of \ac{bler} at the region, where power difference between users is very close to \SI{0}{\deci\bel}. However, as power imbalance between the users is close to each other in conventional \ac{ofdm} \ac{noma}, the performance degrades significantly because power coefficients of the users equate the aggregated signal to the decision boundary. These regions are called as ambiguity region where user's messages are not decoded even with high \ac{snr}. As opposed to conventional power-domain \ac{noma}, the superior region of the user 1 and user 2 in the proposed scheme is roughly below \SI{2}{\deci\bel} and above \SI{-2}{\deci\bel}, respectively. Using forward error correction with soft reconstruction and cancellation removes deep performance losses in the range of certain power differences for \ac{ofdmim}+\ac{ofdm} \ac{noma} scheme. However, conventional power-domain \ac{noma} still has a region where the performance degrades significantly. The proposed waveform-domain \ac{noma} scheme is superior at the region where the power of users is close to each other without having significant performance losses as the power difference between users increases. 
	
\section{Joint Radar-Sensing and Communication Framework with NOMA} \label{sec:jrc_NOMA}
Joint radar-sensing and communication can be actualized using scheduling, data embedded radar waveforms or communication waveforms with radar capabilities including multi or single carrier systems \cite{zheng2019jrcoverview}. Scheduling techniques make efficient use of hardware, but limit the spectral efficiency. Here, we revisit the our conference paper \cite{Sahin2020_MultifunctionalJRC} that proposes the superimposed \ac{fmcw} and \ac{ofdm} waveforms to support the concept of waveform coexistence on \ac{noma}. Proposed architecture separately performs radar-sensing and communication functions using the same radio resources non-orthogonally. 
\subsection{System Model} \label{sec:SystemModel}

\begin{figure}
\subfloat[]{
    \centerline{\includegraphics[width=3.5in]{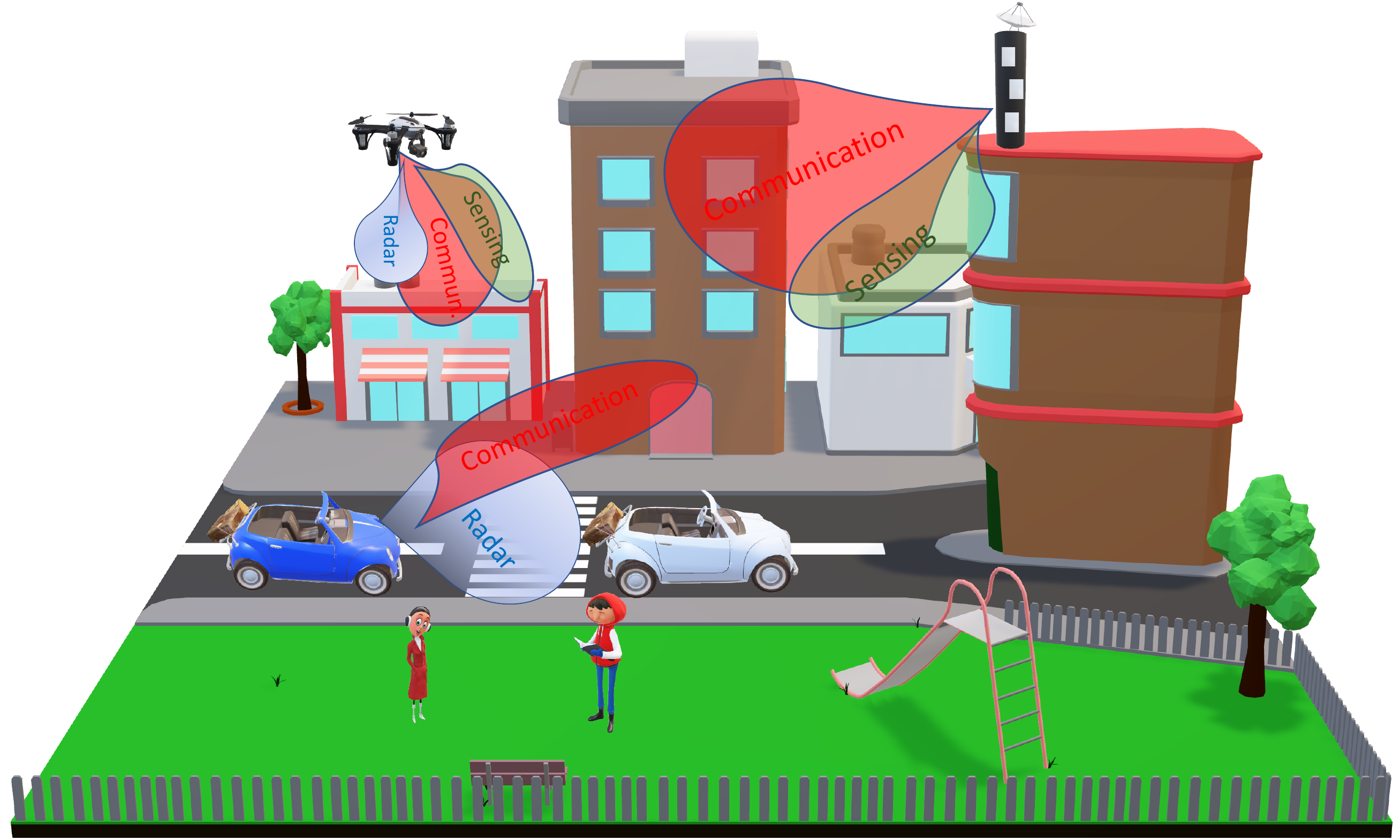}}
    \label{fig:generalRadCom} }
\newline
\subfloat[]{
  \centerline{\includegraphics[width=3.5in]{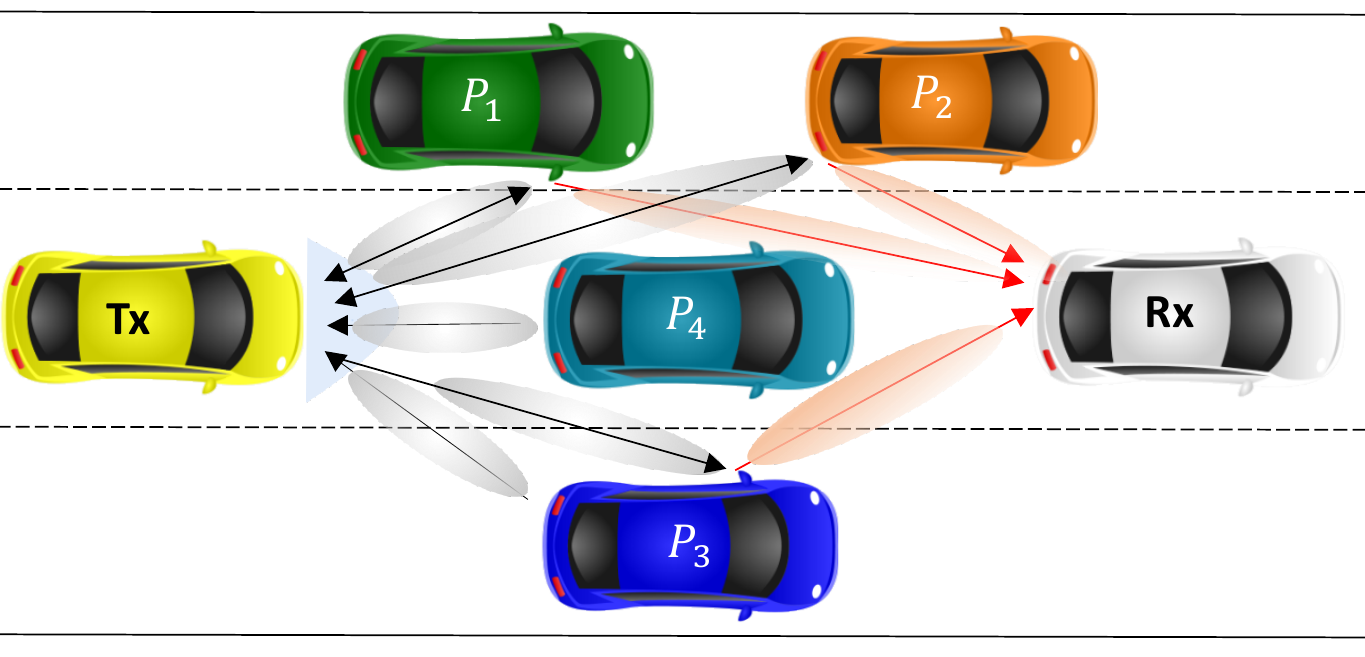}}
\label{fig:V2V_radCom}
}
\caption{System model needing both radar-sensing and communication functionality, (a) general framework, (b) V2V radar-communication.} \label{fig:jrcSystemModel}
\end{figure}

\begin{figure*}
\centerline{\includegraphics[width=7in]{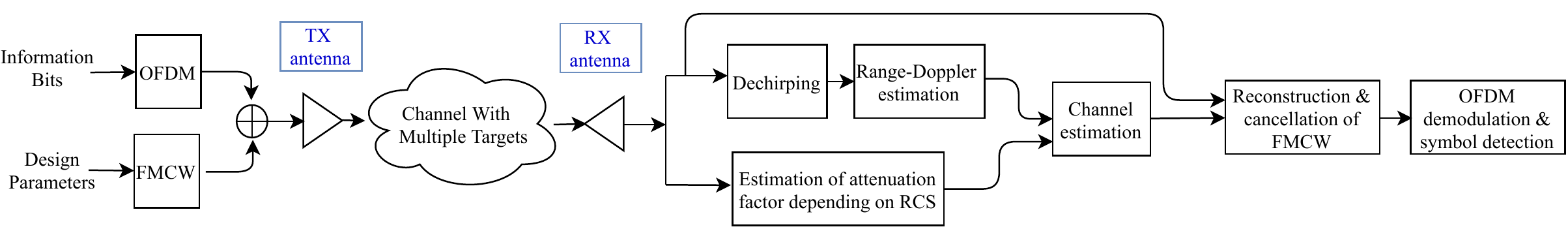}}
\caption{Two users uplink \ac{noma} scheme with \ac{ofdmim}+\ac{ofdm}.\label{fig:jrcScheme}}
\end{figure*}

The V2V scenario is considered as shown in \prettyref{fig:V2V_radCom}, however, the proposed scheme is also applicable for different kinds of use cases needing both radar-sensing and communication activity. For example, more complicated wireless network system is shown in \prettyref{fig:generalRadCom}, including several nodes that function both radar-sensing and communication The general transmission and reception scheme can be seen in \prettyref{fig:jrcScheme} where radar-sensing knowledge obtained from FMCW leveraged to perform channel estimation process in \ac{ofdm} waveform. 

\subsection{Transmission Design}
FMCW consisting of many chirps and OFDM waveforms are utilized for radar-sensing and communication operations, respectively. 
The complex equivalent time-domain representation of one chirp, whose frequency increases linearly across a total bandwidth of $\beta$ Hz during the $\tau$-second is expressed as \cite{richards_2014}
\begin{equation} \label{eq:chirp}
s_\text{chirp}(t) = e^{j\pi\beta t^2/\tau}, \qquad   0\leq t\leq \tau.
\end{equation}
The FMCW waveform consisting of K chirps per frame is
\begin{equation} \label{eq:fmcw}
s_\text{FMCW}(t) = \sqrt{P_\text{FMCW}} \sum_{k=0}^{K-!} s_\text{chirp}(t-k\tau), \qquad   0\leq t\leq T,
\end{equation}
where $T$ and $P_\text{FMCW}$ denote the total duration and power of FMCW waveform, respectively. Number of chirps determines the Doppler resolution when the chirp duration is fixed. On the other hand, the more allocated bandwidth provides the more range resolution in radar-sensing systems.  

Let $\{d_n\}_{n=0}^{N-1}$ be the complex symbols modulated via \ac{qam} drawn from a complex symbol alphabet $\mathbb{S}$. The OFDM signal in time-domain is expressed as 
\begin{equation} \label{eq:ofdm}
s_\text{OFDM}(t) = \sqrt{P_\text{ofdm}} \sum_{n=0}^{N-1} d_n e^{j 2\pi n \Delta f t}, \qquad   0\leq t\leq T_s,
\end{equation}
where $T_s$, $\Delta f$ and $P_{\text{OFDM}}$ denote one OFDM symbol duration, the subcarrier spacing and the OFDM power, respectively.

A cyclic prefix (CP) of length $T_g$ is prepended to each OFDM symbol to keep OFDM subcarriers orthogonal by preventing inter-symbol interference (ISI) across OFDM symbols. CP transforms the linear convolution of the multipath channel to a circular convolution, where one-tap equalization can be used \cite{stuber_2011}. After the CP addition, the $m$th OFDM symbol can be expressed as
\begin{equation} \label{eq:ofdmCP}
\Bar{s}_m(t) = \left\{ \,
\begin{IEEEeqnarraybox}[][c]{l?s}
\IEEEstrut
s_{\text{OFDM}}(t+T_s-T_g) & if $0\leq t\leq T_g$, \\
s_{\text{OFDM}}(t-T_g) & if $T_g < t \leq T_{\text{OFDM}} $,
\IEEEstrut
\end{IEEEeqnarraybox}
\right.
\end{equation}
where $T_{\text{OFDM}} = T_g + T_s$ is the duration of one OFDM symbol after CP addition. 
Having $M$ OFDM symbols in a frame during $T_{\text{sym}} \leq T$, the time domain OFDM signal can be represented as 
\begin{equation} \label{eq:ofdmSymbols}
\Bar{s}_{\text{OFDM}}(t) = \sum_{m=0}^{M-1} \Bar{s}_m(t) \times \text{rect}\left(\frac{t-mT_{\text{OFDM}}}{T_{\text{OFDM}}}\right), \quad 0\leq t\leq T_{\text{sym}}.
\end{equation}

\begin{figure}
\subfloat[]{
    \centerline{\includegraphics[width=3.5in]{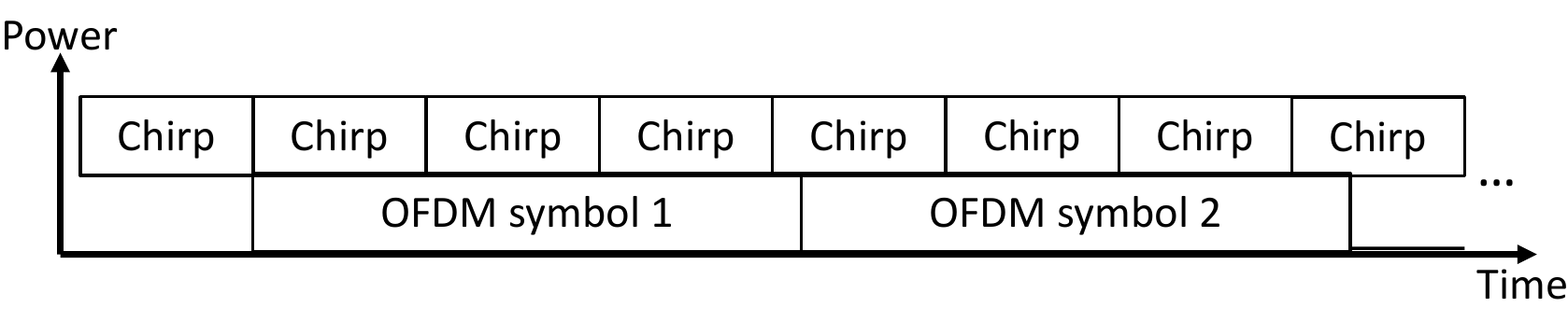}}
    \label{fig:txFrame} }
\newline
\subfloat[]{
  \centerline{\includegraphics[width=3.5in]{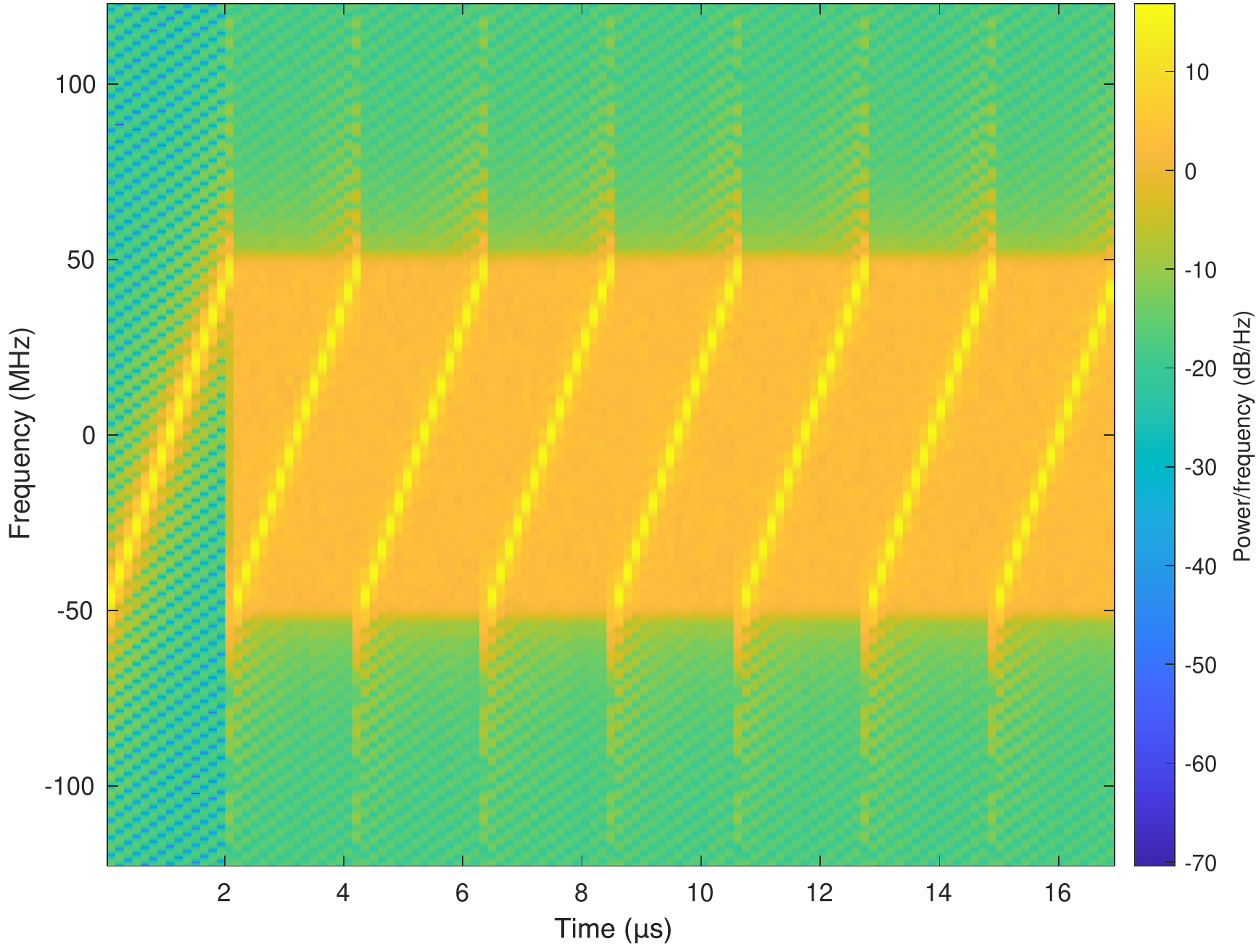}}
\label{fig:txFrameSpectogram}
}
\caption{Proposed transmitted frame design for joint radar and communication functionality, (a) time-power representation, (b) time-frequency representation.}
\end{figure}

In the proposed \ac{noma} scheme, the transmitted frame $s(t)$ includes superimposes both waveforms. To obtain the \ac{rcs} information of the objects in the environment single, single chirp is prepended to the \ac{noma} frame. The final transmit frame for the objectives of multi-functional radar-sensing and communication transmission is designed as follows: 
\begin{equation} \label{eq:transmittedFrame}
{s}(t) = \left\{ \,
\begin{IEEEeqnarraybox}[][c]{l?s}
\IEEEstrut
s_\text{FMCW}(t) & if $0\leq t\leq \tau$, \\
s_\text{FMCW}(t) + \Bar{s}_{\text{OFDM}}(t) & if $\tau < t \leq T $.
\IEEEstrut
\end{IEEEeqnarraybox}
\right.
\end{equation}
The transmitted frame can be seen in \prettyref{fig:txFrame} where it starts with a chirp following the superimposed OFDM symbols and many chirps. The waveforms are superimposed in a way that the allocated bandwidth is the same for each waveform type. The time-frequency illustration of the superimposed signal can be seen in \prettyref{fig:txFrameSpectogram}. It can be realized that the OFDM signal is distributed along with the whole bandwidth, whereas FMCW waveform patterns consecutive pulses whose frequency increases linearly.

Then, the baseband signal $s(t)$ is upconverted to the desired radio frequency (RF) band, where the transmitted passband analog signal becomes 
\begin{equation} \label{eq:passBand}
x(t) = \Re{\{s(t)e^{j (2\pi f_c t + \Bar{\theta})}\}},
\end{equation}
where $\Re{\{.\}}$ denotes the real part of the complex quantity. The notations $f_c$ and $\Bar{\theta}$ are the carrier frequency in which most automotive radars operate in \SI{24}{\giga\hertz} or \SI{76}{\giga\hertz} bands \cite{Ali_2017}, and the initial phase of the transmitted signal, respectively.

\subsection{Channel Effect}


The Doppler shift due to mobility and flight time for the paths reflecting from the targets are as shown in \prettyref{fig:V2V_radCom}. It is assumed that the environment does not change over a coherent transmission time $T$ leading to fixed Doppler shifts and delays. Actually, it is a reasonable assumption that is commonly used in radar literature \cite{richards_2014}.  
With modeling the environment that signal propagates as a linear time-varying channel, the received passband signal is represented as \cite{hlawatsch_matz_2011}
\begin{IEEEeqnarray}{rCl}
r(t) & = & \sum_{p=1}^P \alpha_p \Re{\{x(t-\tau_p)e^{j2\pi(f_c + \psi_p)(t-\tau_p) + j\Bar{\theta} + j \vartheta_p} \}} + n(t), \IEEEeqnarraynumspace  \label{eq:channelModel} 
\end{IEEEeqnarray}
where $\alpha_p$, $\tau_p$ and $\vartheta_p$ are the attenuation factor depending on nonfluctuating radar cross section (RCS), time delay related with the distance between the transmitter to target plus target
to the receiver (bi-static range) and phase error, respectively. The notation $\psi_p = \frac{f_c \upsilon_p}{c}$ is the Doppler frequency associated with the $p$th path depending on relative speed $\upsilon_p$ and the letter $c$ denotes the speed of light. The number $P$ indicates the number of radar targets; in other words, the number of specular scatterer in the environment. Also, $n(t)\sim\mathcal{CN}\left(0,\sigma^2\right)$ denotes the additive white Gaussian noise (AWGN). The attenuation factor $\alpha_p$ is proportional to the large-scale path-loss. Having the path distance $d$ between receiver and transmitter, the large-scale path-loss $G$ is given as 
\begin{equation} \label{eq:pathloss}
	G = \frac{G_{\text{TX}}G_{\text{RX}}\lambda^2}{(4\pi)^2d^\text{PL}}, 
\end{equation}	
where $\text{PL}$ is the path-loss exponent, $G_{\text{TX}}$ and $G_{\text{RX}}$ are the transmit and receive antenna gain, respectively.


\subsection{Multi-functional Reception}\label{sec:reception}
In this section, the receiver scheme for radar-sensing and communication operations is investigated. Since the receiver performs both functions, the knowledge obtained from one process can be leveraged to another to improve the performance.  

\subsubsection{Bi-static Radar Functionality and Channel Estimation}

Down-converting the received passband signal $r(t)$ into baseband and sampling with the frequency of $F_s = N\Delta f$, the discrete-time signal becomes 
\begin{IEEEeqnarray}{rCl} 
\subnumberinglabel{eq:block}
y[n] & = & \sum_{p=1}^P h_p x \left( n/F_s-\tau_p \right) e^{j2\pi n \psi_p / F_s} + w(n), \,\, n \in \mathbb{N}^+, \IEEEeqnarraynumspace  \label{eq:receivedSampled} \\
\noalign{\noindent and
\vspace{2\jot}}
h_p & = & \alpha_p e^{-j 2\pi (f_c + \psi_p)\tau_p + j\Bar{\theta}+j\vartheta_p}, \IEEEeqnarraynumspace  \label{eq:receivedCoeff}
\end{IEEEeqnarray}
where $h_p$ is the complex channel gain of target $p$.
Then, the stretch processing is employed in the discrete domain for the superimposed signal to get delays and Doppler shifts estimations. The processed signal in one chirp time interval can be written as
\begin{equation} \label{eq:chirp}
\Bar{y}[n^\prime] = y[n^\prime] \times e^{-j\pi\beta (n^\prime/F_s)^2/\tau}, \quad n^\prime = 1,2,\ldots, \lfloor \tau F_s \rfloor = N_c,
\end{equation}
and dechirping process is repeated for each chirp time interval. Remember that stretch processing is generally done in time domain with down-conversion. However, here it is assumed that the occupied bandwidth of FMCW is the same as OFDM bandwidth, therefore, the sampling rate $F_s$ for both radar and communication is taken as equal to each other.    
 
By denoting the total chirp count as $K$, a fast-time/slow-time coherent processing interval (CPI) matrix $\mathbf{K} \in \mathbb{C}^{K \times N_c}$ is formed where fast time samples ($l$) are obtained at the rate of $F_s$ from the points on each chirp. Slow-time samples ($k$) are taken from the points on every chirp at the same fast-time sample point. Then this matrix is utilized to perform periodogram based radar processing. The output power of the periodogram at the $m$th Doppler and $n$th range bin is
\begin{IEEEeqnarray}{rCl}
P(m,n) = \frac{1}{KN_c} \Bigg| \underbrace{\sum_{k=0}^{N_c-1} \Bigg( \overbrace{\sum_{l=0}^{K-1} (\mathbf{K})_{k,l} e^{-j2\pi \frac{l m}{K}}}^{N_c \text{ \ac{fft}s of length } K}\Bigg) e^{-j2\pi \frac{kn}{N_c}}}_{K \text{ \ac{fft}s of length } N_c}\Bigg|^2, \IEEEeqnarraynumspace \label{eq:periodogram}
\end{IEEEeqnarray}
sinusoids in $\mathbf{K}$ related to object's distance and velocity lead to peaks in $P(m,n)$. Then certain distance and velocity values can be found from related range and Doppler bin value of peaks. 

Estimation of complex attenuation factor $h_p$ for every  $p$th scatterer (target) is done with the first chirp by the least-square estimation \cite{arslan_bottomley_2001}. It is worth to note that first chirp in the transmitted frame is interference-free as seen in \prettyref{fig:txFrame}. Let the vector $\matr{y_c} \triangleq [y[1], y[2],\ldots,y[\tau F_s]]^\text{T}$ be the samples of the received signal throughout the time $\tau$, the estimated complex attenuation coefficients $\matr{\hat{h}} \triangleq [\hat{h}[1], \hat{h}[2],\ldots,\hat{h}[P]]^\text{T}$ become
\begin{equation} \label{eq:LScoeff}
\matr{\hat{h}} = \argmin_{\matr{h}} (\matr{y_c}-\matr{B}\matr{h})^H(\matr{y_c}-\matr{B}\matr{h}),
\end{equation}
where $\matr{B}$ is a $(\tau F_s - \Bar{n})\times P$ matrix whose rows corresponds to different shifts of the transmitted chirp where shift values are determined according to the estimated range value of the $p$th scatterer (target). The offset value $\Bar{n} \in \mathbb{N}$ depends on the maximum range requirement of the system. Also, the selection of higher value for $\Bar{n}$ decreases fluctuations in the estimation of $\matr{\hat{h}}$ depending on Doppler shifts along with one chirp, whereas the maximum range is reduced.    
By differentiating with respect to $\matr{h}$ and setting the result equal to zero, the least-square estimation of the channel becomes 
\begin{equation} \label{eq:LScoeff}
\matr{\hat{h}} = (\matr{B}^H\matr{B})^{-1}\matr{B}^H\matr{y}.
\end{equation}
Besides the estimation of delays $\tau_p$ and Doppler shifts $\psi_p$, the matrix $\hat{\matr{h}}$ completes the process to recreate the channel matrix $\matr{H}$ with some estimation errors.

\subsubsection{Communication Functionality}
Here, the communication symbols are demodulated using the estimated channel knowledge in the previous section. Let the channel gain of the $k$th sample of the transmitted signal during the reception of the $n$th sample denote as $h_{n,k}$. Also, if the discrete channel convolution matrix along one OFDM symbol duration with $N_{\text{OFDM}}$ samples is shown as $\matr{H} \in \mathbb{C}^{(N_{\text{OFDM}})\times(N_{\text{OFDM}})}$, the element in the $k$th column of the $n$th row of $\matr{H}$ is $h_{n,k}$. 
Firstly, the FMCW sequence is removed from the total received signal by using estimated channel matrix $\matr{\hat{H}}$ as follows: 
\begin{equation} \label{eq:LScoeff}
\matr{y_{\text{OFDM}}} = \matr{y}-\matr{\hat{H}}\matr{s_{\text{FMCW}}},
\end{equation}
where $\matr{y_{\text{OFDM}}}= \left[y_1, y_2, \ldots, y_M\right]$ and $y_m$ is the $m$th OFDM symbol in the received vector $y$. 
The CP addition matrix $\matr{A} \in \mathbb{R}^{N_{\text{OFDM}} \times N}$ is defined as 
\begin{equation}\label{eq:cpAddition}
    A=\left[
\begin{array}{c} 
    \begin{array}{cc}
\matr{0}_{N_g\times N} & \matr{I}_{N_g}
\end{array}
\\  \matr{I_{N_{\text{OFDM}}}}
\end{array}
\right],
\end{equation} 
and the CP removal matrix is generated as $B = \left[\matr{0}_{N\times N_g} \matr{I}_N \right]$ where $N_g$ is the total sample number during CP duration $T_g$.
The matrix $\matr{\Theta} \in \mathbb{C}^{N\times N}$ is the complete channel frequency response (CFR) matrix which equals to 
\begin{equation}
    \matr{\Theta} = \matr{F}_N \matr{B} \matr{\hat{H}} \matr{A} \matr{F}_N^{\text{H}}.
    \label{eq:channel}
\end{equation}
The diagonal components of \prettyref{eq:channel} are the channel coefficients scaling the subcarrier in interest collected in a vector $\boldsymbol{\theta} = \diag{\matr{\Theta}}$ and off-diagonal components are not zero due to Doppler effect from the channel causing ISI. Finally, estimates of data symbols consisting of information bits are obtained as: 
\begin{equation}
    \matr{\hat{d}_m} = \left( \left( \diag{\boldsymbol{\theta} \odot \boldsymbol{\theta}^*} \right)^{-1} (\diag{\theta}^*) \matr{F}_N \matr{B}_K \matr{y}_m \right).
\end{equation}
This equation finalizes the proposed receiver structure without introducing pilots on the OFDM subcarriers to estimate the channel.

\subsection{Numerical Evaluation of the Proposed Scheme}

\begin{table}[!t]
	\renewcommand{\arraystretch}{1}
	\caption{Simulation Parameters}
	\label{tab:simulationParameters}
	\centering
	\begin{tabular}{c||c}
		\hline
		\bfseries Parameter &  \bfseries Value \\
		\hline\hline
		Carrier frequency ($f_c$) & \SI{28}{\giga\hertz}\\
		\hline
		Bandwidth ($\beta$) & \SI{100}{\mega\hertz}\\
		\hline
		Sample Rate ($F_s$) & \SI{122.88}{\mega\hertz}\\
		\hline
		Chirp duration ($\tau$) & \SI{2.4}{\micro\second}\\
		\hline
		FMCW duration ($T$) & \SI{2}{\milli\second}\\
		\hline
		Subcarrier spacing($\Delta f$) & \SI{60}{\kilo\hertz}\\
		\hline
		Number of FFT($N$) & 2048 \\
		\hline
		Number of allocated subcarriers & 1666 \\
		\hline
	\end{tabular}
\end{table}
In this section, the radar-sensing and communication performance of the proposed \ac{noma} scheme is demonstrated. Simulation parameters depending on radar and communication requirements are shown in \prettyref{tab:simulationParameters}. It is assumed that the maximum delay $\tau$ caused by targets is smaller than the CP length of each OFDM symbol. It is assumed that the channel includes 3 targets in the bi-static radar case, as shown in \prettyref{fig:V2V_radCom}. The power delay profile (PDP) of the channel is determined as an exponentially decaying function where the power of channel coefficient is set as $\vert h_p(\gamma) \vert^2 = \eta e^{-\gamma p}$. Let $\eta$ be the normalization factor, $p$ be the target index, $\gamma = 1$ be the decaying factor; and the amplitude of each tap follows Rayleigh distribution. During the simulation, powers of waveforms $P_\text{OFDM}$ and $P_\text{FMCW}$ equal to each other. However, the distribution of power between \ac{fmcw} and \ac{ofdm} can be arranged according to system requirements. Then, it turns out a optimization problem with constraints of data rate and \ac{crlb} of parameter estimation for communication and radar-sensing functionalities, respectively. 

\begin{figure}
\subfloat[]{
    \centerline{\includegraphics[width=0.8\linewidth]{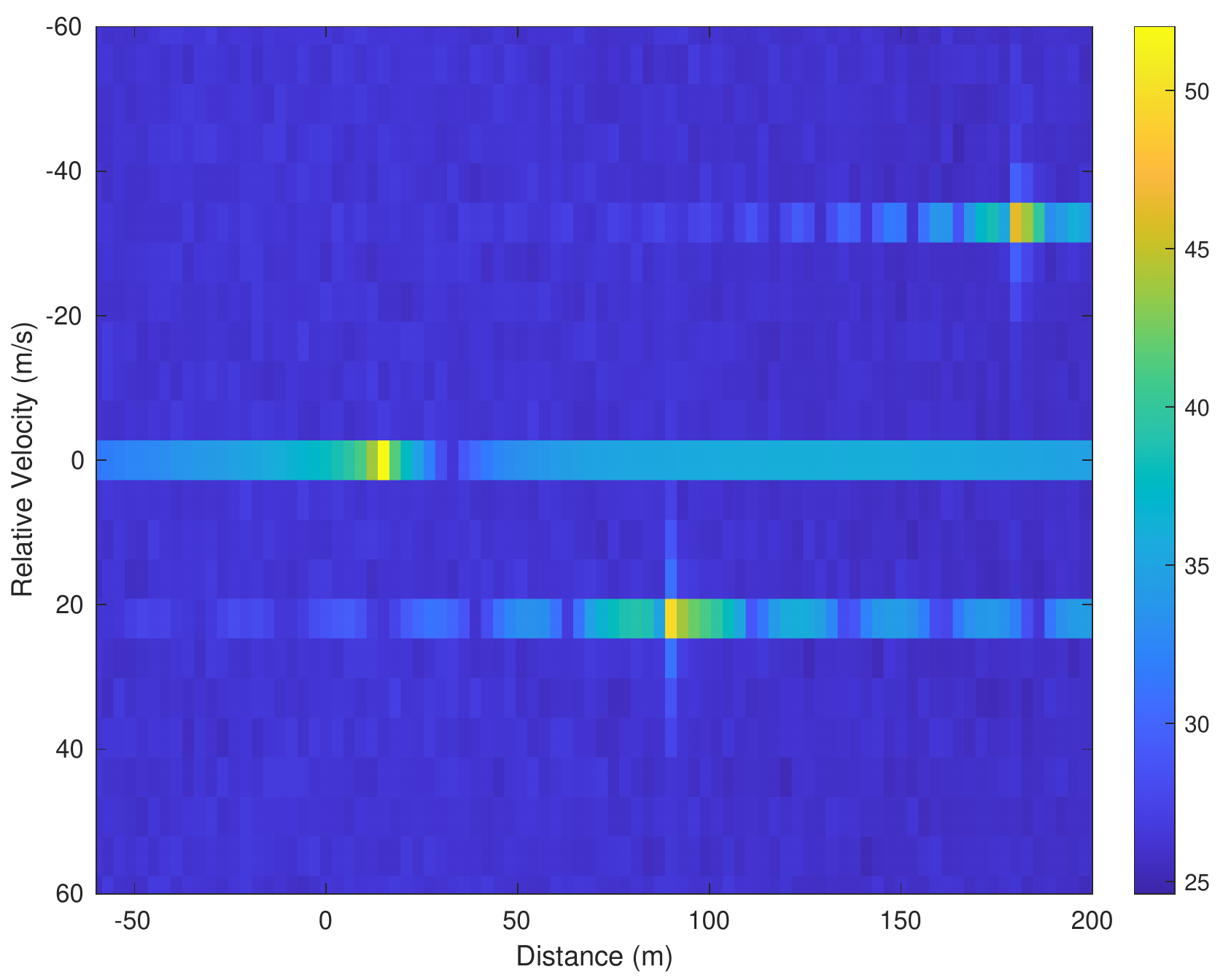}}
    \label{fig:delayDoppler}
}
\newline
\subfloat[]{
  \centerline{\includegraphics[width=0.8\linewidth]{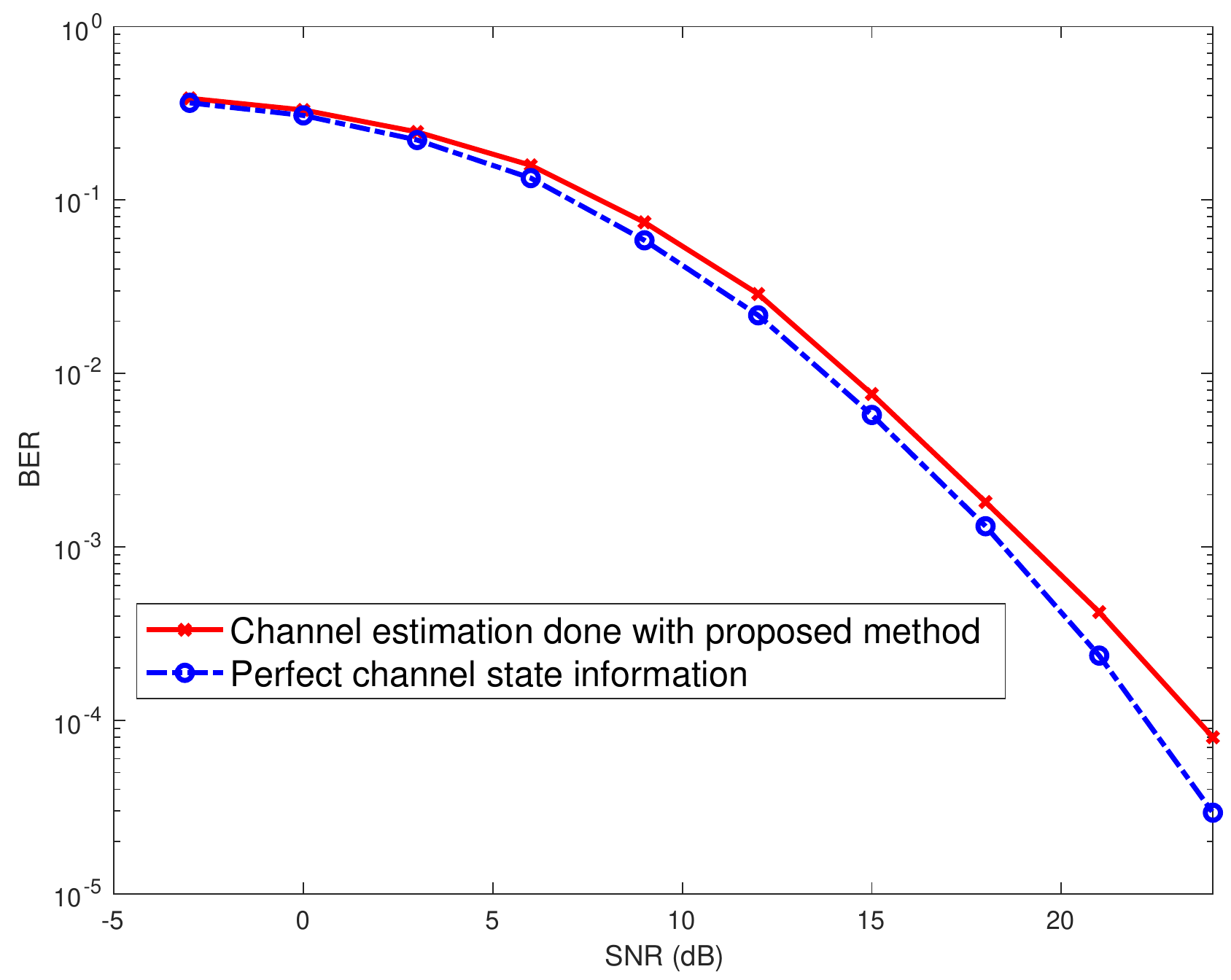}}
\label{fig:berResult}
}
\caption{Performance evaluation of proposed method (a) distance-velocity plot of the targets where SNR equals to $\SI{20}{\deci\bel}$. (b) BER  performance of proposed method.}
\end{figure}

The radar-sensing performance is depicted In \prettyref{fig:delayDoppler} when the \ac{snr} of the superimposed \ac{noma} signal is  $\SI{20}{\deci\bel}$. 
After evaluating the complex attenuation factor $h_p$ denoted in \prettyref{eq:receivedCoeff}, the estimation of channel matrix $\matr{\hat{H}}$ is created by using the obtained values of Doppler shifts and delays which is done previously via FMCW waveform. Finally, this channel estimation is used to demodulate communication symbols in the OFDM waveform. 

The BER performance of the OFDM waveform, where channel estimation knowledge is leveraged from the FMCW waveform, can be seen in \prettyref{fig:berResult}. The information bits are encoded via convolutional codes with interleaving to get rid of the deep fading effect of the channel. The proposed \ac{noma} scheme is compared with the presence of perfect CSI. It means the FMCW waveform is totally removed from the superimposed signal without affecting the OFDM waveform and \ac{ofdm} signal is demodulated with perfect channel estimation. It can be seen that performance degradation of the proposed \ac{noma} scheme is only \SI{0.6}{\deci\bel} at the target BER of 1\%, without requiring any pilot symbols over the OFDM symbols. 
It is worth noting that the complex attenuation factors for each target are estimated using the first chirp, while delays and Doppler shifts are estimated using the FMCW waveform. To sum up, non-orthogonally coexistence of two different waveforms has good sensing accuracy with minimal degradation to communication performance due to slightly lower \ac{snr} as a result of overlapping with the \ac{fmcw} chirps. 

\section{Conclusion} \label{sec:conclusion}
This paper aims to introduce the concept called application-based waveform-domain coexistence on \ac{noma} to meet wide variety of applications proposed in 5G, 6G and beyond wireless networks. The concept emanates from the waveform-domain \ac{noma} principle proposed by the authors where different waveforms are superimposed over the available radio resources. Two main applications of the concept is introduced throughout the paper which are the power-balanced \ac{noma} and joint radar-sensing and communication based on waveform domain \ac{noma}. These approaches clearly indicate the use cases of the concept, however, variation of waveforms regarding applications of wireless networks can be extended. Since the coexistence of different waveforms serves the need of flexibility considering applications and use cases that future wireless systems offers, it is likely that researchers pay attention to improve practicability of the proposed \ac{noma} concept.       
\bibliographystyle{IEEEtran}
\bibliography{open}

\begin{IEEEbiographynophoto}{Mehmet Mert {\c{S}}ahin} (S'20) received the B.S. degree from Bilkent University, Ankara, Turkey, in 2019. He worked at Aselsan Inc. as a wireless communication design engineer in 2019. He is currently working toward the M.S degree at University of South Florida, Tampa, FL, USA. His research interest include waveform design, wireless communication, joint radar-sensing and communication.
\end{IEEEbiographynophoto}

\begin{IEEEbiographynophoto}{H{\"u}seyin Arslan} (S’95–M’98–SM’04–F’15) received the B.S. degree from Middle East Technical University, Ankara, Turkey, in 1992, and the M.S. and Ph.D. degrees from Southern Methodist University, Dallas, TX, USA, in 1994 and 1998, respectively. From 1998 to 2002, he was with the Research Group, Ericsson Inc., NC, USA, where he was involved with several projects related to 2G and 3G wireless communication systems. Since 2002, he has been with the Electrical Engineering Department, University of South Florida, Tampa, FL, USA. He has also been the Dean of the College of Engineering and Natural Sciences, Istanbul Medipol University, since 2014. He was a parttime Consultant for various companies and institutions, including Anritsu Company, Morgan Hill, CA, USA, and The Scientific and Technological Research Council of Turkey (T{\"{U}}B{\.{I}}TAK). His research interests include physical layer security, mmWave communications, small cells, multicarrier wireless technologies, co-existence issues on heterogeneous networks, aeronautical (high-altitude platform) communications, and in vivo channel modeling.
\end{IEEEbiographynophoto}

\end{document}